\documentclass{osa-article}

\journal{osac}

\articletype{Research Article}

\usepackage{lineno}

\newcommand \red[1]{#1}  

\begin{document}

\title{Mode-coupling effects in an optically-injected dual-wavelength laser}

\author{SHAHAB ABDOLLAHI,\authormark{1,*} PABLO MARIN-PALOMO,\authormark{1} and MARTIN VIRTE\authormark{1}}

\address{\authormark{1}Brussels Photonics Team (B-PHOT), Vrije Universiteit Brussel, Pleinlaan 2, 1050 Brussel, Belgium}

\email{\authormark{*}mohammadshahab.abdollahi@vub.be} 



\begin{abstract}
Lasers designed to emit at multiple and controllable modes, or multi-wavelength lasers, have the potential to become key building blocks for future microwave photonic technologies. While many interesting schemes relying on optical injection have been proposed, the nonlinear mode coupling between different modes of a multi-wavelength laser and their dynamical behavior under optical injection remains vastly unexplored. Here, we experimentally and numerically study the effect of optical injection around the suppressed mode of a dual-wavelength laser and the resulting interactions with the dominant mode. We highlight a wavelength shift of the dominant mode triggered by injection locking of the suppressed mode and report a strong impact of the mode suppression ratio on the locking range. Finally, we show numerically that the cross-coupling parameter between the two modes might have a key role in this effect.
\end{abstract}

\section{Introduction}
Semiconductor lasers subject to optical injection display a wide range of dynamical behavior, including stable locking, periodic oscillations, and chaos \cite{Wieczorek2005}. Stable locking is achieved when the injection is strong enough and the detuning – between the wavelength of the master laser and the free-running slave laser – is small enough. Within the locking range, the frequency and phase of the slave laser are locked to those of the master laser. Injection locking has been extensively used to improve laser performance by, e.g., reducing the linewidth \cite{Schunk1986}, suppressing mode hopping and partition noise \cite{Iwashita1982}, or increasing the modulation bandwidth \cite{Simpson1995}. Outside the locking range, different types of dynamics emerge which have been used in various applications such as THz signal generation, chaotic communication, and random bit generation \cite{Wu2017,Liu2001,Li2012}.\\
While the research on optical injection of semiconductor lasers has been mainly carried out with single-mode lasers, multi-wavelength lasers under optical injection are gaining interest over the last years as they may be key enablers of new applications such as all-optical memory \cite{Osborne2009}, reconfigurable photonic microwave generation \cite{Carpintero2018}, and all-optical signal processing \cite{Desmet2020}. In contrast to single-mode lasers, the dynamical behavior of multi-wavelength lasers is strongly influenced by the mode competition occurring through the carrier dynamics in the active medium, and the impact of this competition on the laser dynamics remains to be fully clarified. \\
\red{The self- and cross-saturation induced by spectral effects, such as spectral hole burning, were found to determine the stability of a simultaneous dual-wavelength emission \cite{Agrawal1987}. They depend on the wavelength separation and it was thus predicted that simultaneous dual-wavelength emission would be impossible in quantum well lasers for a mode separation smaller than 0.5 to 1.6~THz \cite{Agrawal1987, Chusseau2013}. Successful experimental demonstration with a 480~GHz separation; however, called for further studies \cite{Osborne2007}. Spatial effects, on the other hand, proved to be important to explain anti-phase dynamics between modes \cite{Masoller2005} and more recently chaos in microlasers \cite{Ma2022}. Yet, their investigations require more complex modelling such as travelling wave models \cite{Homar1996, Serrat2006} or the inclusion of carrier gratings \cite{Viktorov2000, Lenstra2014}. In this case, the coupling doesn't depend on the wavelength separation but rather on the overlap between the cavity eigenmodes \cite{Viktorov2000}.}\\
When considering the effect of optical injection, this coupling between the different laser modes can naturally be expected to have a major impact on dynamical behavior. However, while the case of dual-state emitting quantum dot lasers has attracted considerable attention at least in part due to the particularly complex carrier dynamics \cite{Kelleher2021}, the case of quantum-well multi-wavelength lasers seems to have been left aside – perhaps due to their limited availability – and only a few works tackled this problem \cite{Heinricht2011,Osborneee2009,Osborne2012}. 
\red{Still, in the case of quantum dot or quantum cascade lasers in which multi-wavelength emission typically involves different energy levels, the coupling mechanism could be expected to be significantly different  \cite{Geiser2010, Virte2013a, Chusseau2013}, though similarities have also been reported \cite{Virte2016c}.} \\
In this work, we numerically and experimentally analyze the effect of nonlinear mode coupling in a dual-wavelength laser under optical injection. We focus mainly on the stable locking regime and consider a configuration where one mode is significantly suppressed while the other is dominant. We inject light at a wavelength corresponding to the suppressed mode and analyze the evolution of the locking range for different power balances between the injected and un-injected mode, i.e. the suppressed and dominant mode respectively when no injection is considered. We report a counter-intuitive dependence of the locking range on the power of the suppressed mode: locking appears to be facilitated when the suppressed mode has a higher output power, i.e. the suppression ratio compared to the dominant mode is lower. In addition, within the locking range of the injected mode, we report a detuning-dependent wavelength shift of the un-injected mode which increases with the injection strength and with the power of the suppressed mode, signaling a strong mode coupling in these laser structures. Numerically, we identify the cross-coupling parameter as being particularly relevant. We believe that these results bring new important insight into the mode coupling and competition taking place in multi-wavelength lasers and motivate further investigations focusing on these aspects.

\section{Experimental setup}

\begin{figure}[tb]
\centering\includegraphics[width=\linewidth]{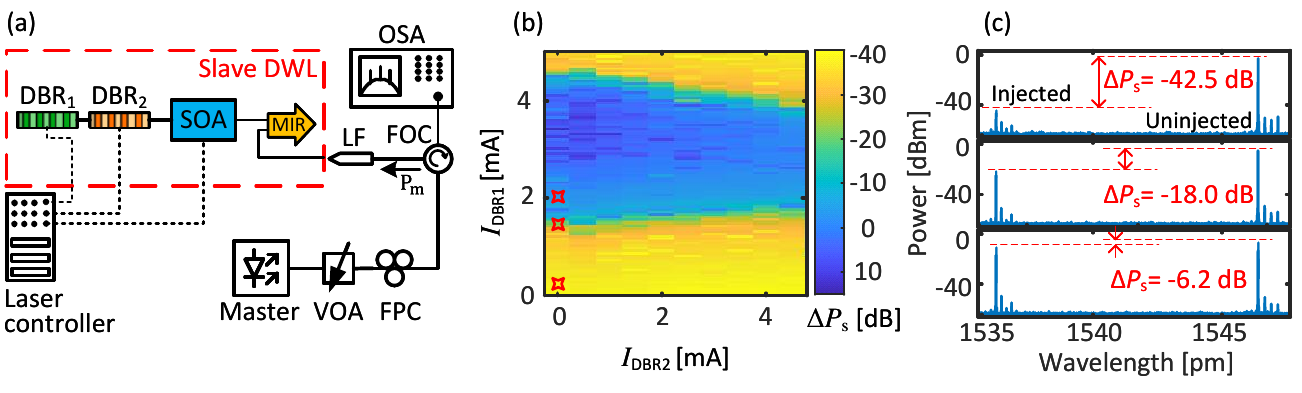}
\caption{(a) Experimental setup: the light from a master laser is routed towards the DWL through a fiber optic circulator (FOC) and a lensed fiber (LF). The injection strength is adjusted by using a variable optical attenuator (VOA). A fiber polarization controller (FPC) is used to adjust the polarization of the injection. A high-resolution optical spectrum analyzer (OSA) is used for measurements. (b) Evolution of the output power ratio $\Delta P_s$ as a function of the current applied to DBR$_1$  and DBR$_2$. The red stars show the three configurations for which the optical spectra are shown in panel (c). (c) Optical spectra of the DWL for the highlighted DBR current configurations.}
\label{fig:Fig1}
\end{figure}

The dual-wavelength lasers (DWL) investigated in this work were fabricated on the InP generic foundry platform of SMART Photonics \cite{Smit2014}, and its structure is schematically described in Fig.~\ref{fig:Fig1}(a). The design uses the standard building blocks of the SMART Photonics library. The gain medium consists of a semiconductor optical amplifier (SOA) with a length of 500~$\mu$m. The laser cavity is formed by a broadband two-port \red{multimode interference reflector (MIR) \cite{Kleijn2013}} on one side and on the other side, two distributed Bragg reflectors (DBR) arranged sequentially which act as wavelength selective elements. The pitch of the DBRs has been set to obtain a spectral separation of 10~nm between both modes. In the investigated structure, the laser emits at $\lambda_1\approx1536.7$~nm and at $\lambda_2\approx1547.6$~nm. DBR$_1$ has a 3-dB bandwidth of 1.08~nm while DBR$_2$ has a bandwidth of 1.47~nm. The parameters of each DBR have been optimized using Lumerical Interconnect simulations \cite{Lumerical} to obtain similar gain levels for both modes of the DWL. With this design, distinct cavities are defined for each wavelength, thus leading to a different separation between longitudinal modes; 31.9~GHz and 41.2~GHz, for wavelengths $\lambda_1$ and $\lambda_2$ respectively. The same laser structure has already been used and its emission properties have been partially characterized in previous works \cite{Pawlusss2019,Pawlus2019}. 
The chip is electrically packaged with all-metal pads being wafer bonded to PCB boards. In addition, it has been glued on a Peltier element including a thermistor. During our experiments, the temperature of the chip is set to 22~ºC. To couple the light out of the photonic integrated circuit, we use a lensed fiber followed by an optical spectrum analyzer (APEX, AP2083, resolution down to 5 MHz / 40 fm) to record the spectrum and monitor the total output power. For a given injection current $I_{SOA}$, the power balance, i.e. the ratio in emitted power of each mode of the DWL, can be controlled by tuning DBR$_1$ and DBR$_2$. By increasing the current applied to the DBR structures, the reflectivity spectrum shifts toward shorter wavelengths while its 3-dB bandwidth and maximum reflectivity decrease slightly \cite{Robbe}. This approach gives a reasonable control of the power balance between the two wavelength emission processes, as can be seen in Fig.~\ref{fig:Fig1}(b) showing the evolution of the power ratios $\Delta P_s = P_{s_1} / P_{s_2}$  between the power, $P_{s_1}$, of the suppressed mode and the power, $P_{s_2}$, of the dominant mode when the current sent to DBR$_1$ and DBR$_2$ is varied. For the record, the subscript ‘s’ stands for “slave laser” as opposed to the master laser output power which will be denoted by an ‘m’ subscript. This evolution is measured for a fixed injection current $I_{SOA}= 60$~mA which is about three times the threshold current of 21~mA. With these parameters, a wide range of $\Delta P_s$ values is achieved. Fig.~\ref{fig:Fig1}(c) depicts the optical spectra for the three DBR current configurations identified by red crosses on the map of the panel (b) and for which power ratios of $\Delta P_s = -42.5 $~dB, -18.0~dB, and -6.2~dB are reported. It should be noted that the total output power of the DWL measured after the lens fiber is $P_s\approx-4.3$~dBm, and remains constant in all tested configurations. Several longitudinal modes can be spotted in the optical spectra but they all remain largely suppressed during our investigations. 

Optical injection is realized by sending the light from a tunable high-quality master laser (Keysight N7776C) into the cavity of the slave DWL via a fiber optic circulator as schematically described in Fig.~\ref{fig:Fig1}(a). A variable optical attenuator is used to adjust the injection strength and a fiber polarization controller to match the polarization of the injected light to that of the DWL. We define the injection strength as the square root of the power of injected light. Because we focus here on the effect of optical injection of the suppressed mode, we define the detuning $\Delta f=f_m -f_{s_1}$ as the frequency difference between the master laser frequency $f_m$ and the free-running frequency of the suppressed mode $f_{s_1}$. 

\section{Identification and classification of partial and full locking}

In contrast to single-mode lasers, we must distinguish in DWL lasers two distinct locking regimes \cite{Heinricht2011}. When the mode under injection is locked to the master laser while the un-injected mode is suppressed, the laser is in the standard locking regime. However, another regime is possible: when the DWL still emits from the un-injected mode while the mode under injection is locked, we denote this regime as partial locking. In this section, we detail how we classify these two regimes experimentally. 

\begin{figure}[tb]
\centering
\includegraphics[width=\linewidth]{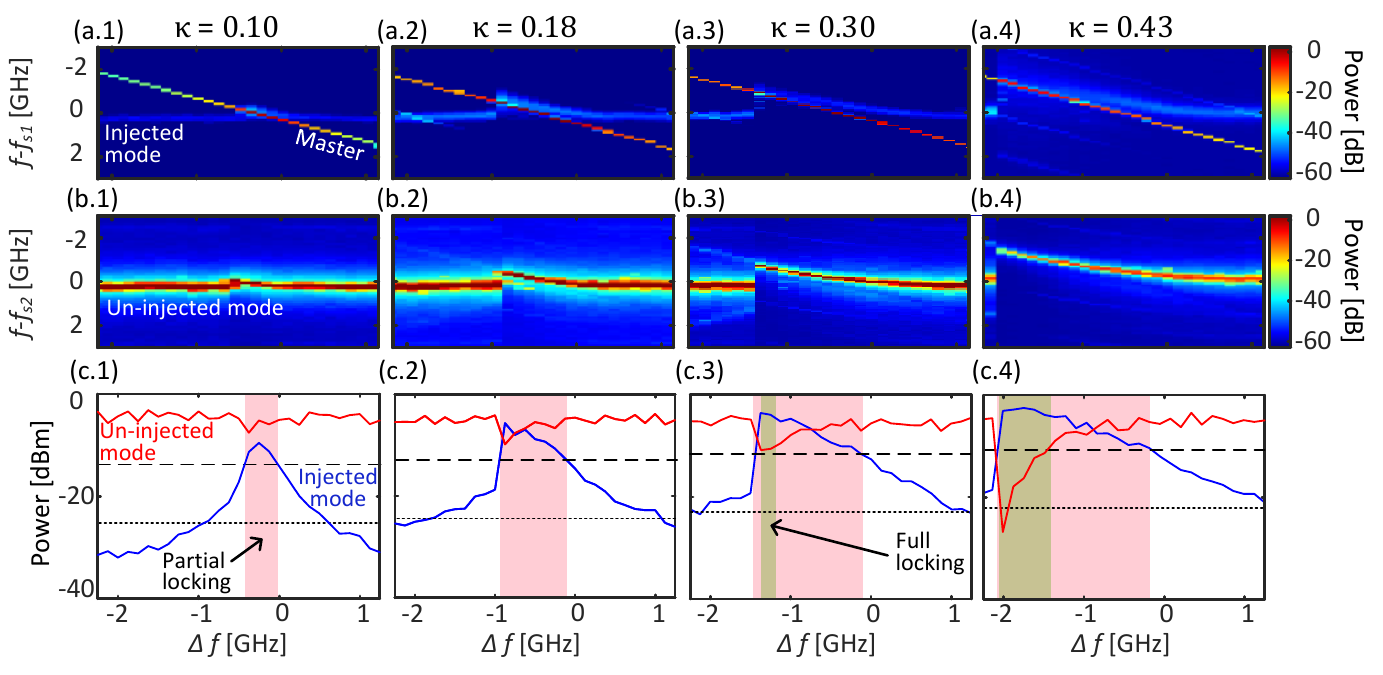}
\caption{Spectral evolution of the DWL modes when the suppressed mode is subject to optical injection for $\Delta P_s=-42.5$~dB while the detuning is changed from -2.3~GHz to 1.3~GHz, for four different $\kappa$ values, from left to right: 0.10, 0.18, 0.30, and 0.43 identified by numbers (1), (2), (3) and (4) respectively. (Top row, a) Spectral evolution of the injected mode $\lambda_1$ with respect to free-running frequency $f_{s_1}$. (Middle row, b) Spectral evolution of the un-injected mode $\lambda_2$ with respect to free-running frequency $f_{s_2}$. (Bottom row, c) Evolution of the peak power around the injected (blue) and un-injected (red) modes. The dotted line represents the sum of the power of the free-running injected mode and master laser. The dashed line represents the 20~dB threshold used to identify partial locking as discussed in the text. The red shaded region indicates the partial locking, and the green shaded regions indicate the full locking range where the laser solely emits at the frequency of the master laser and the un-injected/dominant mode is strongly suppressed.}
\label{fig:Fig2}
\end{figure}

In Fig.~\ref{fig:Fig2}, we show the spectral evolution of the injected/suppressed and un-injected/dominant mode for a power ratio $\Delta P_s=-42.5$~dB when changing the detuning $\Delta f$ from -2.3~GHz to 1.3~GHz for four different injection strengths from $\kappa=0.1$ to 0.43. In Fig.~\ref{fig:Fig2}(a), both master and slave signals are visible in all four cases (a.1), (a.2), (a.3), and (a.4). While locking of the injected mode can be observed in all cases, for an injection strength above $\kappa>0.18$ we also observe a relatively large detuning range where the master laser is strongly influencing the frequency of the injected mode. Interestingly, when looking at the optical spectra of the un-injected mode in the second row, i.e. Fig.~\ref{fig:Fig2}(b), we see that this mode exhibits a similar redshift of its frequency with a similar trend to the one experienced by the injected mode. This is particularly visible when the injection strength is increased and the locking range is extended. We associate this frequency shift with the strong nonlinear coupling of the DWL modes under optical injection.

To further analyze the respective evolution of the injected and un-injected modes, we measure the peak power for each mode at each detuning step, as shown in the four panels of Fig.~\ref{fig:Fig2}(c), the bottom row. Here, we estimate the power of each mode by integrating the optical spectra in a range of $\pm{2.5}$~GHz around the mode frequency before injection, i.e. $f_{s_1}$ and $f_{s_2}$ respectively. For the injected/suppressed mode, with a suppression ratio of $\Delta P_s=-42.5$~dB, the power of the master signal is significantly more powerful than the injected mode. Yet, when locking occurs, we can see a clear increase of the output power of approximately 20~dB, see the blue lines in Fig.~\ref{fig:Fig2}(c.1), (c.2), (c.3), and (c.4). Although this is, in itself, a strong indication of injection locking \cite{Hui1991}, we remark that for a weak injection $\kappa=0.10$, the un-injected mode is still emitting strongly with power levels higher than that of the injected mode. These features, therefore, correspond to what we will define as the partial locking regime. For stronger injection, however, the power of the un-injected mode is experiencing a significant drop. For instance, for $\kappa=0.43$, the power of the un-injected mode is suppressed by approximately 20~dB and the laser emits almost only at the frequency of the master laser. This feature is also consistent with the standard injection locking state which we will refer to as the “full locking” regime as opposed to the “partial locking” discussed above.

In practice, to classify experimentally a given behavior as partial or full locking, we use these different figures of merit and set discriminating thresholds. Although this is, to a certain extent, rather arbitrary and likely imperfect, it gives us a robust, objective, and systematic classification method. First, to identify partial locking, we measure the output power around the injected mode and compare it with the power of the mode in a free-running configuration, i.e. without injection, added to the power of the injected signal, i.e. power of the master laser $P_m$. The latter sum is constant for a given injection strength $\kappa$ and the suppression ratio $\Delta P_s$. Its value is represented in Fig.~\ref{fig:Fig2}(c) by the horizontal dotted line. It can also be seen that this value naturally increases with the injection strength $\kappa$. To identify locking, we then set a threshold chosen heuristically at 20~dB above this dotted line, as shown in Fig.~\ref{fig:Fig2}(c) as the horizontal dashed line. When the measured power is above this threshold, we consider the injected mode to be locked to the master laser if, in addition, the side-mode suppression ratio mode is at least 30~dB. In this case, the side-mode suppression ratio is of course not considering emission around the other wavelength but only a limited bandwidth around the injected mode \cite{Hui1991}. Second, to identify if the locking is partial or total, we compare the power of both injected and un-injected modes. When the power of the injected mode is at least 10~dB higher than the power of the un-injected mode, we consider the DWL to be fully locked to the master laser. Otherwise, if the power difference is less than 10~dB or the un-injected mode is still the dominant one, then we consider the DWL to be partially locked. With this classification method, we can identify the full and partial locking range – highlighted by the green and red shaded area in each panel of Fig.~\ref{fig:Fig2}(c), respectively. Similar to the reports on injection in single-mode lasers, we observe a shift of the locking range towards smaller detuning when increasing the injection strength, which is attributed to the amplitude-phase coupling \cite{Zhang2019}. 

\section{Evolution of the locking range with variations of the suppression ratio}

\begin{figure}[tb]
\centering\includegraphics[width=\linewidth]{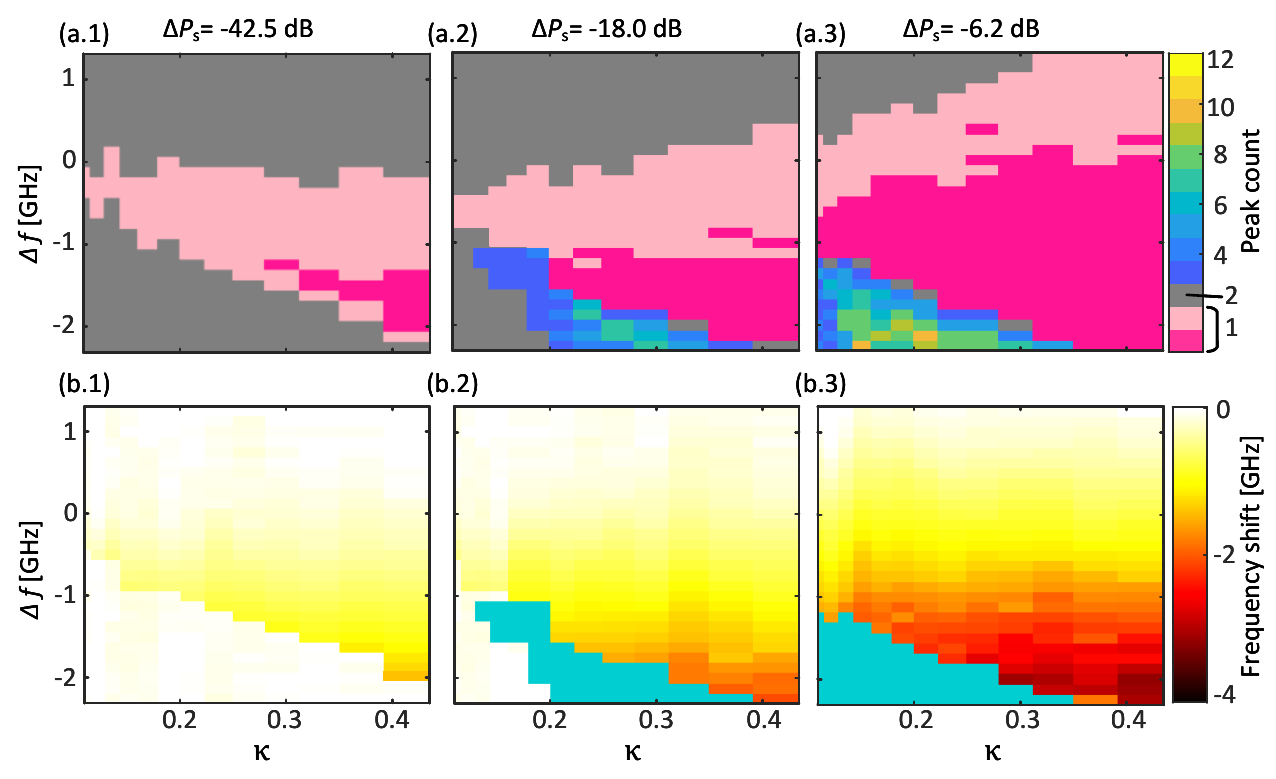}
\caption{Stability maps of the injected mode and frequency shift experienced by the un-injected mode for three different power ratios, $\Delta P_s$ of -42.5~dB (1, left), -18.0~dB (2, middle) and -6.2~dB (3, right). (Top, a) Detuning-injection strength maps for the injected mode. The dark pink and light pink colors illustrate the full and partial locking, respectively. The gray-shaded area indicates the detection of two peaks; thus, no locking is observed, but no dynamics are detected. Higher peak counts are associated with different types of complex dynamics. (b) Frequency shift experienced by the dominant/un-injected mode with respect to its free-running frequency, i.e. without optical injection of the suppressed mode. The blue-shaded area indicates the regions where dynamics are detected, and the data are thus discarded.}
\label{fig:Fig3}
\end{figure}

In this section, we describe the impact that the suppression ratio, $\Delta P_s$, between the injected and un-injected mode has on both the frequency shift experienced by the un-injected mode and the locking range of the DWL. We perform the same measurement as in the previous section, i.e. we inject the suppressed mode and vary the detuning $\Delta f$ from -2.3~GHz to 1.3~GHz, but we repeat it systematically for increasing values of the injection strength $\kappa$ and classify the dynamical state of the laser based on the optical spectrum of the injected mode, i.e. the suppressed mode of our DWL. In  Fig.~\ref{fig:Fig3}(a.1), (a.2), and (a.3), we show these detuning-injection strength maps for three different output power ratios between the two modes: $\Delta P_s=-42.5$~dB, $\Delta P_s=-18.0$~dB, and $\Delta P_s=-6.2$~dB. In each map, we identify the partial and full locking range along with additional complex dynamics featuring broad optical spectra with multiple peaks. Locking is identified by a peak count of one, i.e. only one peak is present in the optical spectrum of the injected mode, i.e. the injected mode is locked to the master laser. This includes both partial and full locking, which, as discussed in the previous section, we distinguish based on the relative output power of each mode. In contrast, a peak count of two represents cases where locking does not occur and both slave and master signals are present in the optical spectrum. Then, higher peak counts indicate the emergence of more complex dynamics such as periodic oscillations or chaos. These states are characterized by broad optical spectra typically comprising several peaks, but, since the precise classification of these dynamics is out of the scope of this work, we only indicate here the number of detected peaks without further details. 

In Fig.~\ref{fig:Fig3}(a.1), (a.2), and (a.3), irrespective of the power ratio between the two modes of the DWL, we observe, as expected, that the locking range broadens when increasing the injection strength. Interestingly, this is true for both partial and full locking. When comparing the three panels, we can clearly notice a significant effect on the suppression ratio $\Delta P_s$ on the locking range: a lower suppression of the mode leads to a large extension of the locking region. In other words, for a given injection strength – estimated with respect to the square root of the power of the injected light – the locking range is extended when the injected mode is emitting at higher power. \red{With a single mode laser subject to optical injection, a more powerful slave laser with a fixed injected power would lead to a decrease of the locking range.} Our observations therefore seems rather counter-intuitive as, from the perspective of the injected mode, the injection strength is, in practice, lower. Our observation, therefore, suggests that the dominant mode might have, in this context, a stabilizing effect on the DWL.

Next, we focus on the evolution of the un-injected/dominant mode and analyze the frequency shift it experiences. Fig.~\ref{fig:Fig3}(b.1), (b.2), and (b.3) show the frequency shift experienced by the un-injected mode while changing the detuning and injection strength applied to the suppressed mode for the three different suppression ratios considered. These values are specified with respect to the mode frequency when no optical injection is applied. In all three cases, we can observe that the un-injected mode experiences a slight redshift whose magnitude depends on both the detuning and injection strength. \red{A similar shift has already been reported in a quantum dot laser but between two longitudinal modes from the same energy level \cite{Hurtado2013}}. When $\Delta P_s$ is increased, both its magnitude and the detuning range in which it appears, increase. The frequency shift clearly occurs when the injected mode is locked. Although it occurs for both partial and full locking regimes, it is significantly larger in the latter. While on the positive detuning side, we observe a smooth transition with the unlocked behavior, a sharp change is observed on the negative detuning side. Complex dynamics appear at the detuning values for which similar dynamical behavior is also observed around the injected mode, see the blue-shaded area in Fig.~\ref{fig:Fig3}(b.2) and (b.3). 

\begin{figure}[tb]
\centering\includegraphics[width=\linewidth]{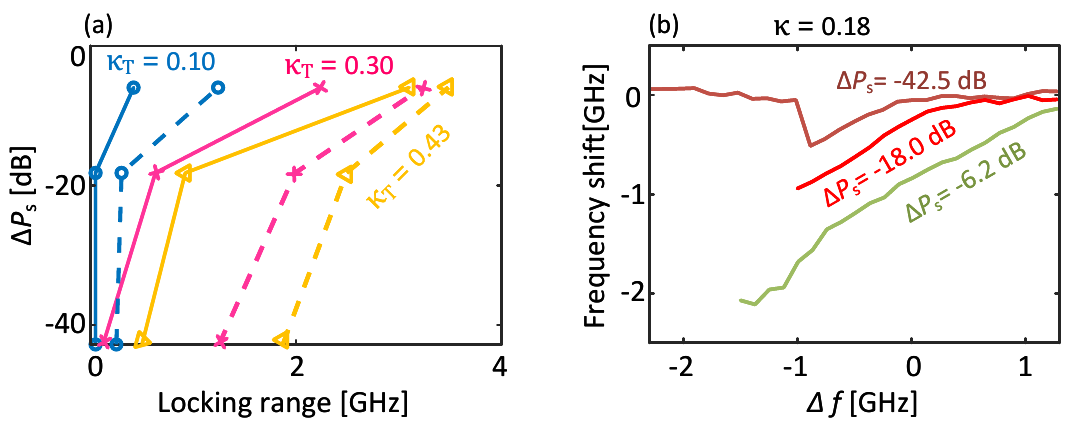}
\caption{Influence of the power ratio, $\Delta P_s$, on the locking range of the injected mode and frequency shift of the un-injected mode. (a) Locking range of the injected mode for an injection strength of $\kappa=0.10$ (blue), $\kappa=0.30$ (pink), and $\kappa=0.43$ (yellow). Solid lines correspond to the full locking range, dashed lines to the full+partial locking range. (b) Frequency shift of the un-injected mode as a function of the detuning for $\Delta P_s=-42.5$~dB (brown), -18~dB (red) and -6.2~dB (green) at a given injection strength of $\kappa=0.18$. For $\Delta P_s=-18.0$~dB and $\Delta P_s=-6.2$~dB, the lines are discontinued when complex dynamics appear on the negative detuning side. }
\label{fig:fig4}
\end{figure}

The suppression ratio $\Delta P_s$ plays an important role in both the locking range and the frequency shift of the un-injected mode. To better analyze its effect, we analyze previous data for different suppression ratios but at a fixed injection strength. In Fig.~\ref{fig:fig4}(a), we compare the evolution of full and full+partial locking ranges at three different injection strengths for increasing suppression ratios. We thus confirm unambiguously that both the locking ranges increase with $\Delta P_s$ regardless of its detailed definition (full or partial). Interestingly, the gap between full and full+partial locking does not appear to be fixed and is even largely reduced at high injection strength and low suppression ($\Delta P_s=-6.2$~dB). Obviously, this feature seems to be attributable to the coupling mechanism between the two modes. Though numerical models provide a good qualitative agreement, as discussed in the next section, the detailed mechanism of this coupling and cross-mode stabilization would require further investigations. Fig.~\ref{fig:fig4}(b) illustrates the detuning-dependent frequency shift of the un-injected mode at three different power ratios. Though already visible in the mapping, this plot clearly shows that increasing $\Delta P_s$ leads to an increase in both the frequency shift magnitude and the detuning bandwidth in which the frequency shift is observed. The slope of the shift curve appears to be identical for all three $\Delta P_s$ configuration and quite close to 1. Since the frequency shift of the un-injected/dominant mode appears while the injected/suppressed mode is locked, we can conclude that the wavelength difference between the two modes of the DWL is preserved despite these shifts. 

\section{Numerical investigations}

In this section, we investigate numerically the locking range of the injected mode and the detuning-dependent frequency shift experienced by the un-injected mode. The model employed for this theoretical analysis is based on the multi-mode extension of the well-known single-mode Lang-Kobayashi equation \cite{Koryukin2004,Viktorov2000}. The equations for the complex fields of each mode, $E_1$ and $E_2$, are given by
\begin{equation}
\frac{dE_1}{dt} =(1+i\alpha)(g_1 N_1-\frac{1-g_1}{2})E_1+\kappa_T e^{i\Delta t},
\end{equation}
\begin{equation}
\frac{dE_2}{dt} =(1+i\alpha)(g_2 N_2-\frac{1-g_2}{2})E_2.
\end{equation}
Compared to \cite{Viktorov2000}, we have removed the term corresponding to the optical feedback but have added an injection term, $\kappa_T e^{i\Delta t}$ to Eq. (1) to account for the optical injection, where $\Delta$ is the normalized detuning and $\kappa_T$ is the injection rate. It should be noted that the injection strength $\kappa$, defined in the previous experimental sections, and injection rate $\kappa_T$ used here cannot be quantitatively compared directly. The reason is two-fold: first, in the experiment, we do not have access to the coupling losses from the lens fiber to the laser cavity on the chip; second, the equations used here are normalized by the photon lifetime as described in \cite{Koryukin2004}. In Eqs. (1) and (2), $\alpha$ represents the linewidth enhancement factor, $g_{1,2}$, the normalized gain coefficients, and $N_{1,2}$ are the carrier densities. Their evolution is described by:
\begin{equation}
\tau_s\frac{dN_1}{dt} =\eta-N_1-(1+2N_1)(g_1|E_1|^2+g_2\beta|E_2|^2),
\end{equation}
\begin{equation}
\tau_s\frac{dN_2}{dt} =\eta-N_2-(1+2N_2)(g_1\beta|E_1|^2+g_2|E_2|^2).
\end{equation}
with $\tau_s$ the normalized carrier lifetime, and $\beta$ the cross-saturation parameter which takes values between 0 and 1. A value of $\beta=0$ describes two decoupled carrier pools whereas $\beta=1$ corresponds to one single carrier pool for both wavelengths.

In this work, unless stated otherwise, we use the following parameter values. The linewidth enhancement factor is set to $\alpha=3$. The pump parameter $\eta$ corresponds to a normalized laser injection current $J$ so that $\eta={J}/{J_{th}}-1$, with  $J_{th}$ being the threshold current. Here, we use $\eta=2$, corresponding to an injection current equal to three times the laser threshold consistent with the conditions of our experimental work. We set the cross-saturation parameter $\beta=0.9$, and the normalized carrier lifetime to $\tau_s=1000$. To tune the suppression ratio $\Delta P_s$ between the two modes of the DWL, we fix the gain of the second mode at $g_2=1$, and adjust the modal gain $g_1$ of mode $E_1$ accordingly to achieve the desired suppression ratio without optical injection. \red{We also consider a standard noise term, not shown, in both field equations with a spontaneous emission factor of $\beta_{sp}=10^{-10}$}. Since the equations are normalized in time with respect to the photon lifetime, all parameters are dimensionless. 

\begin{figure}[tb]
\centering\includegraphics[width=\linewidth]{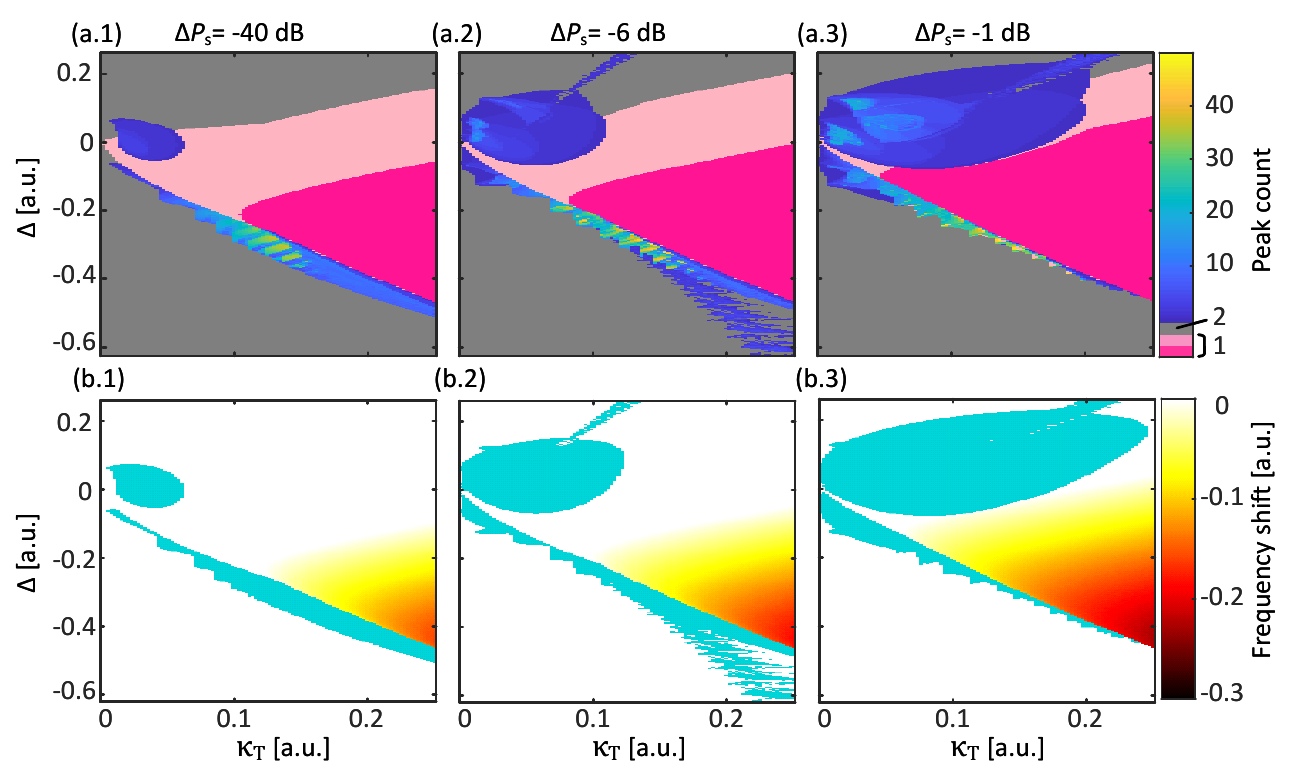}
\caption{Numerical stability maps of the injected mode and frequency shift experienced by the un-injected mode for three different power ratios, $\Delta P_s$ of -40~dB (1, left), -6~dB (2, middle) and -1~dB (3, right). (Top, a) Detuning-injection strength maps for the injected mode. The dark pink and light pink colors illustrate the full and partial locking, respectively. The gray-shaded area indicates the detection of two peaks; thus, no locking is observed, but no dynamics are detected. Higher peak counts are associated with different types of complex dynamics. (b) Frequency shift experienced by the dominant/un-injected mode with respect to its free-running frequency, i.e. without optical injection of the suppressed mode. The blue-shaded area indicates the regions where dynamics are detected, and the data are thus discarded.}
\label{fig 5}
\end{figure}

To reproduce experimental measurements, we sweep the detuning from $\Delta=-0.6$ to 0.2 for increasing the injection rate from $\kappa_T=0$ to 0.25 for three different values of the suppression $\Delta P_s=-40$~dB, -6~dB, and -1~dB. The gain coefficients associated with the mode under injection is varied to achieve the desired $\Delta P_s$ values. $g_1=0.9190$ to achieve $\Delta P_s=-40$~dB, $g_1=0.9234$ to achieve $\Delta P_s=-6$~dB, and $g_1=0.987$ to achieve $\Delta P_s=-1$~dB. We thus obtain the stability maps shown in Fig.~\ref{fig 5}(a.1), (a.2), and (a.3). To identify the partial and full locking range, the same conditions as those described in Sections~2, and 3 are applied. The color code is identical to the one used in Fig.~\ref{fig:Fig3}. The light and dark pink colors indicate both partial and full locking regions respectively, where only one peak is detected in the optical spectrum of the injected mode. The grey color depicts the regions where two peaks are detected reflecting the fact that both master and slave lasers are present in the optical spectrum and no locking occurs. Higher peak counts represent the complex dynamical behavior in which more than 2 peaks are detected. By increasing $\kappa_T$ or $\Delta P_s$, both partial and full locking regions broaden, i.e. the higher the injection strength or the power balance ($\Delta P_s$) the wider the locking range. This feature coincides well with our experimental observations. \red{We did not observe any sign of hysteresis or multi-stability in the partial or full locking regions}. However, we also observe large regions of complex dynamics on the positive detuning side which is clearly visible for all three suppression ratios configurations which we have not observed experimentally. Although this dynamical region is consistent with the standard behavior of a single-mode laser subject to optical injection \cite{Wieczorek2005}, it was clearly missing in our experimental measurements. Experimentally, we only achieved a $\Delta P_s$ up to -6~dB and, in this case, the dynamical region appears for relatively low injection rate which can be reached mostly before the full locking. Since in Fig.~\ref{fig:Fig3}(a.3), full locking is observed for all injection strength values considered, we might suppose that these complex dynamics might occur for weaker injections that were out of reach for our experimental setup. Nevertheless, a more detailed analysis of these cases might be required to clarify this point. 

In Fig.~\ref{fig 5}(b), we monitor the spectral evolution of the un-injected mode while the suppressed mode is under injection. As in the experiment, the frequency of the un-injected mode experiences a frequency shift toward lower frequency values. Both the magnitude and detuning range in which the frequency shift occurs, are affected by the power ratio between the injected and un-injected mode. The detuning range in which the frequency shift occurs aligns perfectly with the full locking range of the injected mode. In addition, we detect complex dynamical behavior – shown with blue shaded color – which coincides with that observed in the optical spectrum of the injected mode.

\begin{figure}[tb]
\centering\includegraphics[width=\linewidth]{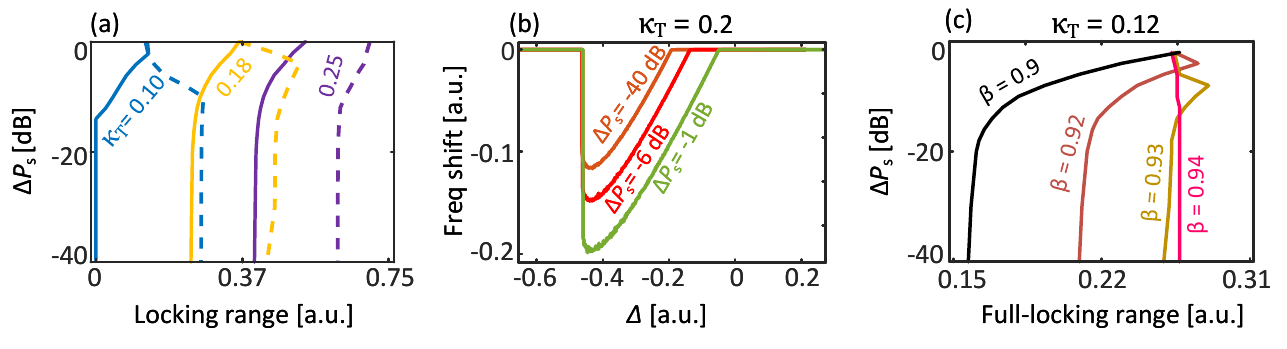}
\caption{Numerical analysis of the impact of the suppression ratio on both the locking bandwidth of the injected mode and frequency shift experienced by the un-injected mode and the influence of the cross-coupling between two modes of the DWL on the full locking bandwidth. (a) The locking bandwidth of the injected mode with respect to the power ratio between injected and un-injected modes at $\kappa_T=0.10$ (blue), $\kappa_T=0.18$ (yellow), and $\kappa_T=0.25$ (purple). The full locking is shown by the solid lines and the dashed lines correspond to the full+partial locking. (b) Frequency shift of the un-injected mode for an injection rate $\kappa_T=0.2$ as a function of detuning at $\Delta P_s=-40$~dB (brown), -6~dB (red) and -1~dB (green). (c) Full locking range as a function of the suppression ratio for an injection rate of $\kappa_T=0.12$ and increasing cross-coupling values $\beta=0.9$ (black), 0.92 (brown), 0.93 (yellow) and 0.94 (pink).}
\label{fig 6}
\end{figure}

To better understand the impact of the power ratio on the locking range of the injected mode, we focus on three different injection strengths, $\kappa_T=0.10$, $\kappa_T=0.18$, and $\kappa_T=0.25$, and identify the full and full+partial locking range as done experimentally. Again, we adjust $g_1$ to tune the suppression ratio $\Delta P_s$, and compute the locking range for each value, see Fig.~\ref{fig 6}(a). We observe that, again, the full locking range increases by increasing $\Delta P_s$. However, for the full+partial locking, a slightly different behavior is observed. The full+partial locking decreases for $\Delta P_s$ higher than 10~dB or 5~dB for $\kappa_T=0.10$ and $\kappa_T=0.18$ respectively. This variation can be connected to the emergence of the complex dynamics region that appears on the positive detuning side of the locking range as seen in Fig.~\ref{fig 5}(a.2) and (a.3). At higher injection rates, however, the same trend can still be observed. 

To investigate the dependence of the frequency shift of the un-injected mode on the suppression ratio between the two modes of the DWL, we set the injection strength to $\kappa_T=0.2$. Next, we retrieve the frequency shift experienced by the un-injected mode at three different power ratios and over the whole detuning range, see Fig.~\ref{fig 6}(b). By reducing the power ratio between the two modes of the DWL, the frequency shift magnitude and the detuning range in which the frequency shift occurs, increase. Overall, we again obtain a good agreement with the experimental observations but also with a few interesting discrepancies. The evolution of the frequency shift appears to be less linear than in the experiment and, even though the slope of the shift is again identical for all $\Delta P_s$ configurations, the slope is not close to one. Thus, meaning that, unlike in the experiment, the injection modifies the wavelength separation between the two modes. This result suggests that the wavelength separation might not be as invariant as one could have supposed based on the experimental results. 

To go further, and because all these features seem to be intrinsically linked with the coupling mechanism between the two modes of the DWL, we numerically analyze the impact of the cross-saturation parameter $\beta$. In Fig.~\ref{fig 6}(c), we plot the full locking range of the injected mode with respect to the suppression ratio $\Delta P_s$ for $\kappa_T=0.12$ and different $\beta$ values. Interestingly, we observe that, by increasing the cross-coupling between the injected and un-injected mode, the dependence of the full locking range on the suppression ratio is drastically impacted to the point of being suppressed for $\beta=0.94$. The cross-saturation between the two modes of the DWL, therefore, seems to have a central role in the response of the DWL to optical injection. 

\section{Summary}
In this work, we have experimentally and numerically investigated the effect of optical injection on the suppressed mode of a dual-wavelength laser. We have highlighted an important dependence of the locking range on the suppression ratio between the two modes of the laser, along with a significant frequency shift of the un-injected mode. We obtained a good agreement between our experimental observations and modeling based on a rather simple rate equation model including a cross-saturation between the two modes. We have highlighted that this cross-saturation parameter might have a leading role in shaping the dynamical behavior of DWL and, in particular, their response to optical injection. \red{The rate equation model including carrier-grating and cross-saturation effects appears to be sufficient to qualitatively reproduce the behavior observed experimentally. Yet, at this stage, we cannot fully dismiss the role of other coupling mechanisms even though this analysis is outside of the scope of this paper.} Further investigations might be required to fully uncover and understand the mode coupling in multi-wavelength lasers.

\begin{backmatter}
\bmsection{Funding}
Research Foundation - Flanders (FWO) (grants 1530318N, G0G0319N), METHUSALEM program of the Flemish Government (Vlaamse Overheid), the European Research Council (grant 948129 COLOR-UP). 

\bmsection{Disclosures}
The authors declare no conflicts of interest.

\bmsection{Data availability} Data underlying the results presented in this paper are not publicly available at this time but may be obtained from the authors upon reasonable request.

\bibliography{sample, 20220712_ExtraRef_OEpaperDWLwInjection_SA}

\begin{thebibliography}{10}
\newcommand{\enquote}[1]{``#1''}

\bibitem{Wieczorek2005}
S.~Wieczorek, B.~Krauskopf, T.~Simpson, and D.~Lenstra, \enquote{The dynamical
  complexity of optically injected semiconductor lasers,}
  {\protect\JournalTitle{Physics Reports}} \textbf{416}, 1--128 (2005).

\bibitem{Schunk1986}
N.~Schunk and K.~Petermann, \enquote{Noise analysis of injection-locked
  semiconductor injection lasers,} {\protect\JournalTitle{IEEE J. Quantum
  Electron}} \textbf{22}, 642--650 (1986).

\bibitem{Iwashita1982}
K.~Iwashita and K.~Nakagawa, \enquote{Suppression of mode partition noise by
  laser diode light injection,} {\protect\JournalTitle{IEEE Transactions on
  Microwave Theory and Techniques}} \textbf{30}, 1657--1662 (1982).

\bibitem{Simpson1995}
T.~B. Simpson, J.~M. Liu, and A.~Gavrielides, \enquote{Bandwidth enhancement
  and broadband noise reduction in injection-locked semiconductor lasers,}
  {\protect\JournalTitle{IEEE Photonics Technol. Lett.}} \textbf{7}, 709--711
  (1995).

\bibitem{Wu2017}
J.-W. Wu, Q.~Qiu, X.-P. Zhang, and Y.~H. Won, \enquote{Simultaneous generation
  of microwave, millimeter-wave, and terahertz photonic signal based on
  two-color semiconductor laser subject to single-beam optical injection,}
  {\protect\JournalTitle{IEEE Journal of Selected Topics in Quantum
  Electronics}} \textbf{23}, 1800108 (2017).

\bibitem{Liu2001}
J.~M. Liu, H.~F. Chen, and S.~Tang, \enquote{Optical-communication systems
  based on chaos in semiconductor lasers,} {\protect\JournalTitle{IEEE Trans
  Circuits Syst}} \textbf{48}, 1475--1483 (2001).

\bibitem{Li2012}
X.~Z. Li and S.~C. Chan, \enquote{Random bit generation using an optically
  injected semiconductor laser in chaos with oversampling,}
  {\protect\JournalTitle{Opt. Lett}} \textbf{37}, 2163--2165 (2012).

\bibitem{Osborne2009}
S.~Osborne, K.~Buckley, A.~Amann, and S.~O'brien, \enquote{All-optical memory
  based on the injection locking bistability of a two-color laser diode,}
  {\protect\JournalTitle{Opt. Express}} \textbf{17}, 6293--6300 (2009).

\bibitem{Carpintero2018}
G.~Carpintero, S.~Hisatake, D.~de~Felipe, R.~Guzman, T.~Nagatsuma, and N.~Keil,
  \enquote{Wireless data transmission at terahertz carrier waves generated from
  a hybrid {InP}-polymer dual tunable {DBR} laser photonic integrated circuit,}
  {\protect\JournalTitle{Scientific Reports}} \textbf{8}, 3018 (2018).

\bibitem{Desmet2020}
R.~Desmet and M.~Virte, \enquote{Laser diodes with modulated optical injection:
  towards a simple signal processing unit?} {\protect\JournalTitle{Journal of
  Physics: Photonics}} \textbf{2}, 025002 (2020).

\bibitem{Agrawal1987}
G.~Agrawal, \enquote{{Gain nonlinearities in semiconductor lasers: Theory and
  application to distributed feedback lasers},} {\protect\JournalTitle{IEEE
  Journal of Quantum Electronics}} \textbf{23}, 860--868 (1987).

\bibitem{Chusseau2013}
L.~Chusseau, F.~Philippe, P.~Viktorovitch, and X.~Letartre, \enquote{{Mode
  competition in a dual-mode quantum-dot semiconductor microlaser},}
  {\protect\JournalTitle{Physical Review A}} \textbf{88}, 015803 (2013).

\bibitem{Osborne2007}
S.~Osborne, S.~O'Brien, K.~Buckley, R.~Fehse, A.~Amann, J.~Patchell, B.~Kelly,
  D.~R. Jones, J.~O'Gorman, and E.~P. O'Reilly, \enquote{Design of single-mode
  and two-color fabry - pérot lasers with patterned refractive index,}
  {\protect\JournalTitle{IEEE Journal of Selected Topics in Quantum
  Electronics}} \textbf{13}, 1157--1163 (2007).

\bibitem{Masoller2005}
C.~Masoller, M.~S. Torre, and P.~Mandel, \enquote{{Antiphase dynamics in
  multimode semiconductor lasers with optical feedback},}
  {\protect\JournalTitle{Physical Review A}} \textbf{71}, 013818 (2005).

\bibitem{Ma2022}
C.-g. Ma, J.-L. Xiao, Z.-X. Xiao, Y.-D. Yang, and Y.-Z. Huang,
  \enquote{{Chaotic microlasers caused by internal mode interaction for random
  number generation},} {\protect\JournalTitle{Light: Science \& Applications}}
  \textbf{11}, 187 (2022).

\bibitem{Homar1996}
M.~Homar, S.~Balle, and M.~S. Miguel, \enquote{Mode competition in a
  fabry-pérot semiconductor laser: travelling wave model with asymmetric
  dynamical gain,} {\protect\JournalTitle{Optics Communications}} \textbf{131},
  380--390 (1996).

\bibitem{Serrat2006}
C.~Serrat and C.~Masoller, \enquote{{Modeling spatial effects in
  multi-longitudinal-mode semiconductor lasers},}
  {\protect\JournalTitle{Physical Review A}} \textbf{73}, 043812 (2006).

\bibitem{Viktorov2000}
E.~A. Viktorov and P.~Mandel, \enquote{Low frequency fluctuations in a
  multimode semiconductor laser with optical feedback,}
  {\protect\JournalTitle{Physical Review Letters}} \textbf{85}, 3157--3160
  (2000).

\bibitem{Lenstra2014}
D.~Lenstra and M.~Yousefi, \enquote{{Rate-equation model for multi-mode
  semiconductor lasers with spatial hole burning},}
  {\protect\JournalTitle{Optics Express}} \textbf{22}, 8143 (2014).

\bibitem{Kelleher2021}
B.~Kelleher, M.~Dillane, and E.~A. Viktorov, \enquote{Optical information
  processing using dual state quantum dot lasers: complexity through
  simplicity,} {\protect\JournalTitle{Light: Science \& Applications}}
  \textbf{10}, 238 (2021).

\bibitem{Heinricht2011}
P.~Heinricht, B.~Wetzel, S.~O’Brien, A.~Amann, and S.~Osborne,
  \enquote{Bistability in an injection locked two color laser with dual
  injection,} {\protect\JournalTitle{Applied Physics Letters}} \textbf{99},
  011104 (2011).

\bibitem{Osborneee2009}
S.~Osborne, A.~Amann, K.~Buckley, G.~Ryan, S.~P. Hegarty, G.~Huyet, and
  S.~O’Brien, \enquote{Antiphase dynamics in a multimode semiconductor laser
  with optical injection,} {\protect\JournalTitle{Physical Review A}}
  \textbf{79}, 023834 (2009).

\bibitem{Osborne2012}
S.~Osborne, P.~Heinricht, N.~Brandonisio, A.~Amann, and S.~O’Brien,
  \enquote{Wavelength switching dynamics of two-colour semiconductor lasers
  with optical injection and feedback,} {\protect\JournalTitle{Semiconductor
  Science and Technology}} \textbf{27}, 094001 (2012).

\bibitem{Geiser2010}
M.~Geiser, C.~Pfl{\"{u}}gl, A.~Belyanin, Q.~J. Wang, N.~Yu, T.~Edamura,
  M.~Yamanishi, H.~Kan, M.~Fischer, A.~Wittmann, J.~Faist, and F.~Capasso,
  \enquote{{Gain competition in dual wavelength quantum cascade lasers},}
  {\protect\JournalTitle{Optics Express}} \textbf{18}, 9900 (2010).

\bibitem{Virte2013a}
M.~Virte, K.~Panajotov, and M.~Sciamanna, \enquote{{Mode competition induced by
  optical feedback in two-color quantum dot lasers},}
  {\protect\JournalTitle{IEEE Journal of Quantum Electronics}} \textbf{49},
  578--585 (2013).

\bibitem{Virte2016c}
M.~Virte, R.~Pawlus, M.~Sciamanna, K.~Panajotov, and S.~Breuer,
  \enquote{{Energy exchange between modes in a multimode two-color quantum dot
  laser with optical feedback},} {\protect\JournalTitle{Optics Letters}}
  \textbf{41}, 3205 (2016).

\bibitem{Smit2014}
M.~Smit and et. al., \enquote{An introduction to {InP}-based generic
  integration technology,} {\protect\JournalTitle{Semiconductor Science and
  Technology}} \textbf{29}, 083001 (2014).

\bibitem{Kleijn2013}
E.~Kleijn, M.~K. Smit, and X.~J. Leijtens, \enquote{Multimode interference
  reflectors: A new class of components for photonic integrated circuits,}
  {\protect\JournalTitle{Journal of Lightwave Technology}} \textbf{31},
  3055--3063 (2013).

\bibitem{Lumerical}
\enquote{Lumerical interconnect, “synopsys, inc. (2022),” 2022. [online].
  available \url{https://www.lumerical.com/products/interconnect},} .

\bibitem{Pawlusss2019}
R.~Pawlus, S.~Breuer, and M.~Virte, \enquote{Compact dual-wavelength lasers on
  {InP} generic foundry platform,} in \emph{2019 Conference on Lasers and
  Electro-Optics Europe \& European Quantum Electronics Conference
  (CLEO/Europe-EQEC),}  (2019), pp. 1--1.

\bibitem{Pawlus2019}
R.~Pawlus, R.~de~Mey, S.~Breuer, and M.~Virte, \enquote{Dual-wavelength lasers
  on generic foundry platform,} {\protect\JournalTitle{ArXiv preprint}} p.
  1911.03533 (2019).

\bibitem{Robbe}
R.~de~Mey, R.~Pawlus, S.~Breuer, and M.~Virte, \enquote{{Characterization and
  wavelength control of DBR-based dual-wavelength lasers},} in
  \emph{Semiconductor Lasers and Laser Dynamics IX,}  vol. 11356 (SPIE, 2020),
  pp. 65 -- 72.

\bibitem{Hui1991}
R.~Hui, A.~D'Ottavi, A.~Mecozzi, and P.~Spano, \enquote{Injection locking in
  distributed feedback semiconductor lasers,} {\protect\JournalTitle{IEEE J.
  Quantum Electron}} \textbf{27}, 1688--1695 (1991).

\bibitem{Zhang2019}
Z.~Zhang, D.~Jung, J.~C. Norman, W.~W. Chow, and J.~E. Bowers,
  \enquote{Linewidth enhancement factor in {InAs/GaAs} quantum dot lasers and
  its implication in isolator-free and narrow linewidth applications,}
  {\protect\JournalTitle{IEEE Journal of Selected Topics in Quantum
  Electronics}} \textbf{25}, 1900509 (2019).

\bibitem{Hurtado2013}
A.~Hurtado, I.~D. Henning, M.~J. Adams, and L.~F. Lester, \enquote{Generation
  of tunable millimeter-wave and {THz} signals with an optically injected
  quantum dot distributed feedback laser,} {\protect\JournalTitle{IEEE
  Photonics Journal}} \textbf{5}, 5900107--5900107 (2013).

\bibitem{Koryukin2004}
I.~V. Koryukin and P.~Mandel, \enquote{Dynamics of semiconductor lasers with
  optical feedback: Comparison of multimode models in the low-frequency
  fluctuation regime,} {\protect\JournalTitle{Phys. Rev. A}} \textbf{70},
  053819 (2004).

\end{thebibliography}

\end{backmatter}






\end{document}